\documentstyle[aps,epsf,prl,multicol]{revtex}
\begin{document}
\vspace{5.0cm}
\draft
\title{Delocalization border and onset of chaos in a model of
quantum computation}

\author{G.P.Berman$^{[a]}$, F.Borgonovi$^{[b,c]}$,
F.M.Izrailev$^{[d]}$, V.I.Tsifrinovich$^{[e]}$ }
\address{
$^{[a]}$Theoretical Division and CNLS, Los Alamos National
Laboratory, Los Alamos, New Mexico 87545\\ $^{[b]}$Dipartimento di
Matematica e Fisica, Universit\`a Cattolica, via Musei 41, 25121
Brescia, Italy  \\ $^{[c]}$ I.N.F.M., Gruppo Collegato di Brescia
and I.N.F.N., Sezione di Pavia   Italy\\ $^{[d]}$Instituto de
Fisica, Universidad Autonoma de Puebla, Apdo. Postal J-48, Puebla
72570, Mexico\\ $^{[e]}$ IDS Department, Polytechnic University, Six Metrotech
Center, Brooklyn, New York 11201\\
}
\date{\today}
\maketitle
\begin{abstract}

We study the properties of spectra and eigenfunctions for a
chain of $1/2- $spins (qubits) in an external time-dependent
magnetic field, and under the conditions of non-selective excitation (when the amplitude of the magnetic field is large). This model is known as a possible candidate for
experimental realization of quantum computation. We present the
theory for finding delocalization transition and show that for the
interaction between nearest qubits, the transition is very
different from that to quantum chaos. We explain this phenomena by
showing that in the considered region of parameters our model is
close to an integrable one. According to a general opinion, the
threshold for the onset of quantum chaos due to the interqubit
interaction decreases with an increase of the number of qubits.
Contrary to this expectation, for a magnetic field with constant
gradient we have found that chaos border does not depend on the
number of qubits. We give analytical estimates which explain this
effect, together with numerical data supporting our analysis.
Random models with long-range interactions are studied as well. In
particular, we show that in this case the delocalization and
quantum chaos borders coincide.
\end{abstract}
\begin{multicols}{2}

\section{Introduction}

In recent years much attention has been paid to the idea of
quantum computation \cite{F86}. The burst of interest to this
subject (see, for example, \cite{S98,BDMT98,N00} and references
therein) is caused by the discovery of fast quantum algorithms for
the factorization of integers \cite {S94} and for the effective
searching of items in a database \cite{G97,CGK98}. These
algorithms demonstrate the effectiveness of quantum computers in
comparison with the classical ones. Nowadays, there are different
projects for the experimental realization of quantum computers, as
well as experimental results with few-qubit systems (see \cite {exper}) and references therein).

Main theoretical suggestions for the experimental implementation
of the quantum computation are based on interacting two-level
systems ({\it qubits} ). It is clear that one of the most
important problems from the viewpoint of the stability of quantum
operations, is a destructive role of different kinds of errors. In
the first line, one should refer to finite temperature effects and
the interaction with an environment \cite{CLSZ95}. However, even
in the case when these effects can be neglected, there are
dynamical effects of the interqubit interaction, which may
influence a quantum computation. On one hand, the interaction
between qubits is necessary for the realization of quantum
computation, on the other hand, it may result in a kind of
destruction of the coherence in the evolution of a system.

The latter subject of the dynamical decoherence is directly
related to the so-called {\it quantum chaos} which is nowadays
widely discussed in application to atoms, nuclei, quantum dots and
other physical systems (see, for example, \cite{chaos} and
references therein). One of the latest developments in the theory
of quantum chaos refers to the interaction between Fermi-particles
in isolated systems. The core of this approach is the perturbation
theory for many-body states, which takes into account a
two-body nature of the interaction. Specifically, it was shown
\cite{AGKL97} that if the two-body random interaction between
particles exceeds some critical value, fast transition to chaos
occurs in the Hilbert space of many-particle states (see also
\cite{A90,FIC96,FI97,all} and reviews \cite {GMW99,var99}).

In dynamical systems such as complex atoms \cite{Ce}, multicharged
ions \cite{ions}, nuclei \cite{nuclei} and spin systems
\cite{Nobel,spins}  quantum chaos gives rise to a very
complicated structure of highly excited states, and to specific
correlations in the energy spectra, described by Random Matrix
Theory  (RMT) (see, for example, \cite{GMW99}). As a result, closed
dynamical systems with relatively small number of interacting
particles can be well described by a statistical approach, see
discussion and references in \cite{I01}.

Recently,  quantum chaos theory has been applied to a simple
model of quantum computer \cite{dima} chosen in the form of $L$
interacting qubits. Numerical data have shown that for a strong
enough interaction between qubits the onset of quantum chaos is
unavoidable. Although for $L=14-16$ the critical value $J_{cr}$
for the quantum chaos border was found to be quite large, with an
increase of $L$ the border decreases as $J_{cr}\sim 1/L$
\cite{dima0,dima}. From the viewpoint of the standard approach for
closed systems of interacting particles, the decrease of chaos
border with an increase of qubits looks generic. This poses the
question of the relevance of quantum chaos to quantum computation
\cite{F00,SSB99}.

In our recent paper \cite{gena} we have studied the errors which appear in
the evolution of 1D Ising nuclear spins in rotating magnetic field. This
model was suggested for an experimental realization of a quantum computer
\cite{ber1,ber2}. The main attention in \cite{gena} has been paid to
the region of parameters, most suitable for the preparation of an
initial many-body state needed for further application of {\it
quantum protocol} (sequence of time-dependent magnetic pulses in a
prescribed algorithm of quantum computation). It was shown that
even for a very large interqubit interaction, the errors turn out
to be very small, thus demonstrating that the influence of quantum
chaos can be neglected.

An analysis of the stationary Hamiltonian describing the system
during a single magnetic pulse has been performed in \cite{qcomp}.
Specifically, the general approach of quantum chaos theory has
been applied, in order to understand the conditions for the onset
of quantum chaos. The model we considered assumed that qubits 
(nuclear spins) are placed in a strong magnetic field with  
constant gradient along the direction of the spin  chain. 
The gradient of the magnetic field provides
a ``labeling'' of qubits. Namely, each spins has different Larmor frequency, $\omega_k$. This allows one to provide a selective addressing to each qubit by applying resonant {\it rf} pulses. The main interest was
in the influence of the magnetic field on the properties of
eigenstates and energy spectra. It was unexpectedly found that the
constant gradient magnetic field gives rise to the independence of
the critical value $J_{cr}$ on the number $L$ of interacting
qubits. This striking phenomena has been explained in \cite
{qcomp} analytically and confirmed numerically, thus giving a new
insight to the problem of quantum chaos in the models of quantum
computers.

In this paper we present the full theory which explains the
properties of energy spectra and many-body states of the model of
Ref. \cite{gena}, together with numerical data obtained in a broad
region of the model parameters. The structure of the paper is as
follows. In the next section we describe the model, discuss the
region of parameters of our interest, and briefly analyze the
structure of the Hamiltonian matrix in the $z$-representation. In
Sect. III we study global properties of the energy spectrum, paying
main attention to the band structure of the spectrum and to the
level spacing distribution $P(s)$ for the central energy band.

Section IV is the core of the paper, here both the delocalization
border and the condition for the onset of quantum chaos are
studied. The consideration has been made by making use of the
mean-field representation which is very convenient from the
theoretical viewpoint. One of two main goals of this section is
that these two borders are very different in the model with
nearest interaction between qubits. Another important result is
that the delocalization border turns out to be independent of the
number of qubits for a gradient magnetic field. Theoretical
estimates obtained in this Section serve as a guiding line to
treat all numerical data.

In Section V we investigate numerically the structure of
eigenstates in the $z$-representation, by relating the data with
the theoretical predictions. Section VI is devoted to some
modifications of the model, namely, we analyze the influence of
randomness in the interqubit interaction. Our main question is how
statistical properties of the system depend on the range of the
interaction between qubits. Specifically, we study random
interaction between all qubits ($A$ -interaction), as well as
between four nearest qubits, by comparing the results with those
obtained for the model with the interaction between two nearest
qubits ($N$-interaction).

General discussion is presented in the last Section VII. One of
the problems we discuss here, is the concept of the
quasi-integrability of our model for the $N$-interaction. We show
that for the region of parameters of our interest, the model is
close to the integrable one. This explains why the delocalization
and chaos borders do not coincide for the $N$-interaction. We also
analyze the role of the magnetic field. In particular, we give
analytical estimates which show that for the homogeneous magnetic
field the delocalization border has generic $L$-dependence
discussed in \cite{dima}. On the other hand, for the magnetic
field with an increasing gradient, analytical estimates predict
that the delocalization border increases with an increase of the
number of qubits.

\section{The model}

The model describes a 1-dimensional chain of $L$ interacting
distinguishable $1/2$-spins in an external magnetic field.
Schematically, these spins ({\it qubits}) can be represented as
follows,

$$
\Uparrow B^z:\quad\uparrow_{L-1}\downarrow_{L-2}...
\uparrow_{1}\uparrow_{0}.
$$

\noindent Here $B^z$ stands for a constant part of magnetic field oriented
in the positive $z$-direction, and each qubit occupies one of two
single-particle states with the energy $1/2$ (position ``up'') or
$-1/2$ (position ``down''). One can see that the total number $N$
of {\it many-body states} which are generated by this chain ({\it
quantum register}), is $N=2^L $.

The dynamics of this model ({\it quantum computer protocol}) is due to a sum
of $p=1,...,P$ time-dependent rectangular pulses of a circular polarized
magnetic field rotating in the $x,y$-plane. Each of the pulses has own
amplitude $b^p_\perp$, frequency $\nu_p$, phase $\varphi_p$, and lasts
during the period $T_p=t_{p+1}-t_p$. Therefore, the total magnetic field
during one pulse can be written as follows,\cite{gena},
\begin{equation}
\label{sp}{\vec B}(t)=[b^p_\perp\cos(\nu_p t+\varphi_p),-b^p_\perp\sin(\nu_p
t+\varphi_p), B^z],
\end{equation}
The Hamiltonian of this system has the form,
\begin{equation}
\begin{array}{ll}
{\cal H}= -\sum\limits^{L-1}_{k=0} (\omega_kI^z_k+2 \sum\limits_{n >
k}J_{k,n} I^z_k I^z_n)- & ~~~~~~~~~ \\
{\frac{{1}}{{2}}}\sum\limits_{p=1}^{P}\Theta_p(t)\Omega_p
\sum\limits_{k=0}^{L-1} \Bigg(e^{-i\nu_p t-i\varphi_p}I^-_k+ e^{i\nu_p
t+i\varphi_p}I^+_k\Bigg), \label{ham00}
\end{array}
\end{equation}
where the ``pulse function" $\Theta_p(t)$ equals $1$ only during the $p$-th
pulse of the length $T_p$. The quantities $J_{k,n}$ stand for the Ising
interaction between two qubits , $\omega_k$ are the frequencies of spin's
precession in the $B^z-$magnetic field, $\Omega_p$ is the Rabi frequency
corresponding to the $p$-th pulse. The operators $I_k^{\pm}$
are defined by the relations $I_k^{\pm}=I^x_k \pm iI^y_k$, and
 $I_k^{x,y,z} = (1/2) \sigma_k^{x,y,z}$, the
latter being the Pauli matrices.

Below we consider the properties of the system during a single
$p$-th pulse. The corresponding Hamiltonian can be written in the
coordinate system which rotates around $z$-axes with the frequency
$\nu_p$. Thus, for the $p$-th
pulse, our model can be reduced to the {\it stationary}
Hamiltonian,
\begin{equation}
\begin{array}{ll}
{\cal H}^{(p)}=-\sum\limits_{k=0}^{L-1} [(\omega_k-\nu_p)I^z_k+
\Omega_p(\cos\varphi_p I^x_k-\sin\varphi_p I^y_k)+ & ~~~~~~~~~ \\
2 \ \sum\limits_{n>k}^{}J_{k,n}I^z_k I^z_n], \label{ham}
\end{array}
\end{equation}
which describes the evolution of the model for $t_p<t\le t_{p+1}$.

The regime of quantum computation corresponds to the following range of
parameters: $\Omega_p\ll J_{k,n}\ll\delta\omega_k\ll\omega_k$, where $%
\delta\omega_k=|\omega_{k+1}-\omega_k|$ \cite{gena} (the so-called {\it %
selective excitation}). In this regime, each pulse acts  
selectively on a chosen qubit exciting a resonant transition. 
The inequality, $\Omega_p\ll J_{k,n}$, provides a separation 
between 
resonant and non-resonant transitions for the same selected qubit. 
The inequality, $J_{k,n}\ll\delta\omega_k$, provides a 
separation of transitions for a given qubit from the 
transitions for neighboring qubits.
 In this paper we consider another
important regime of {\it non-selective excitation} which is defined by the
conditions, $\Omega_p\gg \delta\omega_k \gg J$, see details in \cite{gena}.
This inequality provides the simplest way to prepare a homogeneous
superposition of $2^L$ states needed for implementation of both Shor and
Grover algorithms.

In what follows we assume, for simplicity, $\varphi _p=\pi /2$,
and put $\Omega _p=\Omega $ and $\nu _p=\nu $. Our main interest
is in the nearest neighbor interaction ({\it N-interaction})
between qubits for two different cases, the {\it dynamical} one
when all coupling elements are the same, $ J_{k,n}=J\ \delta
_{n,k+1}$, and the case when all values $J_{k,k+1}$ are random
({\it random model}). However, we will also analyze other cases
with different kinds of interaction and compare results with those
for the $N$-interaction. In contrast to the previously discussed
model \cite{dima} with homogeneous magnetic field, below we
consider the magnetic field which depends on the position of the
$k$-th qubit. Therefore, we assume that the spin frequencies
$\omega _k$ are slightly dependent on $k$ (with $\delta
\omega _k\ll \omega _k$).

For the dynamical $N-$interaction, the Hamiltonian (\ref{ham}) takes the
form,
\begin{equation}
\label{ham0}H=\sum_{k=0}^{L-1} \Big [-\delta_kI^z_k+ \Omega I^y_k\Big]
-2J\sum_{k=0}^{L-2} I^z_k I^z_{k+1}.
\end{equation}
where $\delta_k=\omega_k-\nu$. In the $z$-representation the
Hamiltonian matrix of size $N = 2^L$ is diagonal for $\Omega=0$.
For $\Omega\not=0$ the off-diagonal matrix elements are
$H_{k,n}=i\Omega/2$ for $n>k$, and $H_{nk}=H^*_{kn}$. When
calculating the matrix elements of the Hamiltonian (\ref{ham0}) we
have used the standard rules in order to find the action of the
operators $I^z_k$ and $I^y_k$ on the states $|k\rangle$ and
$|n\rangle$, $$
I^z_k|...0_k...\rangle={\frac{{1}}{{2}}}|...0_k...\rangle,\quad
I^z_k|...1_k...\rangle=-{\frac{{1}}{{2}}}|...1_k...\rangle, $$ $$
I^y_k|...0_k...\rangle={\frac{{i}}{{2}}}|...1_k...\rangle,\quad
I^y_k|...1_k...\rangle=-{\frac{{i}}{{2}}}|...0_k...\rangle. $$

The matrix turns out to be very sparse, and it has quite specific
structure (see Fig.1) in the basis which is reordered according to
an increase of the number $s$ written in the binary
representation, $s=i_{L-1}, ... ,i_1,i_0$ (with $i_s=0$ or $1$,
depending on whether a single-particle state of $i-$th qubits is
the ground state or the excited state). In what follows, we call
this representation as the $z-$representation.

\begin{figure}
\epsfxsize 6cm
\epsfbox{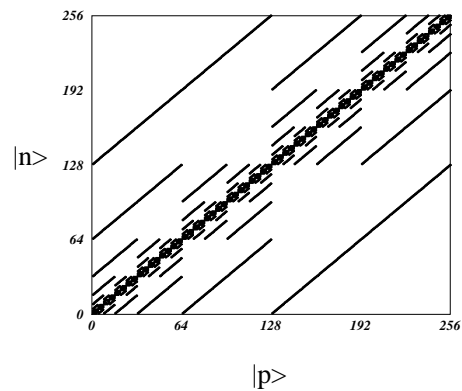}
\narrowtext
\caption{
Structure of the Hamiltonian (\ref{ham0}) for $L=8$ in the $z$
-representation. Dots stand for the matrix elements different
from zero. }
\label{mat}
\end{figure}

\section{Global properties of the energy spectrum}

For the further analysis, it is important to understand the global
structure of the energy spectrum. In what follows, we concentrate
our attention to the case when the magnetic field has a constant
gradient along the chain of qubits, $w_k=w_0+ak$ with $a>0$. Other
cases will be briefly discussed in Sect. VII.

\subsection{Band structure}

\begin{figure}
\epsfxsize 7cm
\epsfbox{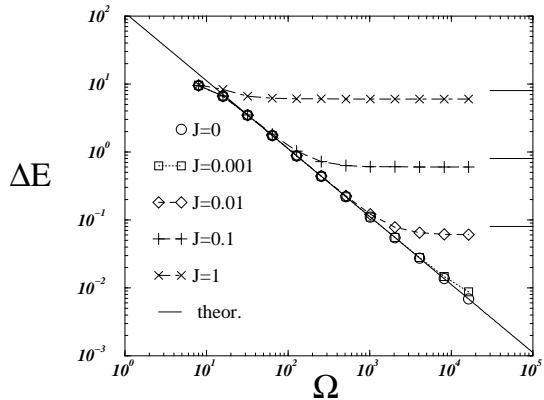}
\narrowtext
\caption{
Dependence of the central bandwidth $\Delta E$ on $\Omega$ for
different values of $J$. The data are shown for $L=10$,
$\omega_k=\omega_0+k$, $\nu=\omega_0$ and $a=1$. The full straight
line is the theoretical expression for $J=0$. The horizontal lines
on the right hand side of the figure correspond to the analytical
expression for $(\Delta E)_2$ for the case of $J>J_{s}$, (see in
the text). }
\label{bband}
\end{figure}

Without the interaction between qubits, $J=0$, the energy spectrum
of the model (\ref{ham0}) consists of $L+1$ {\it bands} of finite
width for $a\neq 0 $, separated by big gaps of size $\Omega
\gg\omega_k$. In Ref.\cite {gena,qcomp} it was numerically found
that the width $\Delta E(\Omega, J=0)$ of the central band
decreases with an increase of $\Omega$ as $A_L /\Omega$. Our
analytical estimates show that for $L$ even, the bandwidth is
given by the relation $(\Delta E)_1 = L^2 a^2 (L-1)/8\Omega $,
(see details in Sect. IV). This dependence also occurs for a
relatively weak interaction $ J\neq 0$. However, when the
interaction exceeds some critical value $J_{s}$, the band widths
turn out to be practically independent on $\Omega$, see the data
for the central band in Fig. \ref{bband}.

The bandwidth $(\Delta E)_2$ for the interaction strength $J$
larger than the critical value $J_{s}$ can be also estimated
analytically as $(\Delta E)_2 = (L-2) a J$ (see Sect. IV). The
correspondence between the analytical estimate and numerical data
was found to be quite good. If the band-width $\Delta E $ is
larger than $(\Delta E)_2$, the first expression $(\Delta E)_1$
for $\Delta E (\Omega) $ dominates. Contrary, if the band-width
$(\Delta E)_2$ defined by the interaction $J$ is larger, it
determines the actual band-width $\Delta E$ which is independent
of $\Omega$.

\begin{figure}
\epsfxsize 6cm
\epsfbox{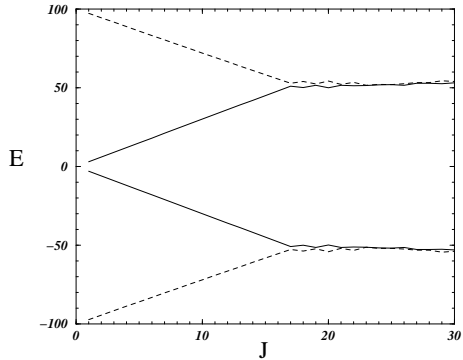}
\narrowtext
\caption{
Energy bands as a function of $J$. Only the central band (full
line) and its neighbors (dashed lines) are shown, thus
demonstrating the band overlapping for a relatively large
interaction. The parameters are $L=10$, $\Omega=100$,
$\omega_k=\omega_0+k$, $\nu=\omega_0$ and $a=1$. }
\label{width}
\end{figure}

One should stress that the above consideration is valid for the
case when the bands are not overlapped. One can expect that for
sufficiently strong interaction between the qubits, the band
structure of the energy spectrum disappears. The overlapping of
the central band with two other bands is shown in Fig.\ref{width}
where the edges of the central and the nearest bands are plotted
in dependence on the interaction $J$ for the fixed value $\Omega =
100$. One can see that for $J>J_b\approx 15$ the bands are
overlapped, therefore, a change in the properties of the
system is naturally expected. The critical value $J_b$ for the
overlapping of the bands is estimated in Sect. IV as well.

\subsection{Level spacing distribution}

Let us now analyze the distribution $P(s)$ of spacings $s$ between
nearest-neighbor energy levels inside the central energy band
(note that $s$ should be normalized to the mean spacing between
levels). This quantity is often used in the theory of {\it Quantum
Chaos} as a detector of chaotic properties of a system.
Specifically, for systems with regular motion in the classical
limit, the distribution $P(s)$ is generically close to the Poisson
(apart from 1-dimensional systems where $P(s)$ is highly
non-generic and can be of any form). In the other limit case of a
completely chaotic motion, in the corresponding quantum systems
the distribution $P(s)$ has the so-called {\it Wigner-Dyson} (WD)
form which is characterized by the level repulsion for small
spacings, $s \ll1$ ($P(s)\sim s, s^2, s^4$, depending on the
symmetry of a system, see, e.g. \cite{GMW99}).

\begin{figure}
\epsfxsize 7cm
\epsfbox{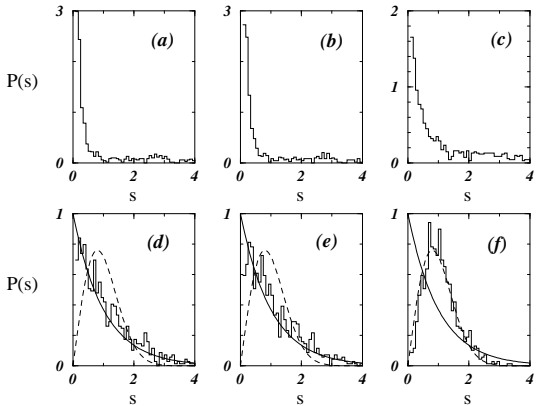}
\narrowtext
\caption{
Level spacing distribution $P(s)$ in dependence on the interaction
$J=0, 0.0002, 0.1,1, 10, 100 $ denoted by ($a,b,c,d,e,f$)
respectively. Other parameters are $L=12, \Omega=100$ ,
$\omega_k=\omega_0+k$, $k=1,...,L$, $\omega_0=100$, $\nu=\omega_0$
and $a=1$. Full curve is the Poisson, dashed curve corresponds to
the Wigner-Dyson distribution. Eigenvalues are taken from the
central energy band only. }
\label{pds}
\end{figure}

Numerical data for $P(s)$ for different values of the interaction
strength $J$,  summarized in Fig.\ref{pds}, manifest the transition
to the WD-distribution. Note that for small values of $J$ (see
Fig.\ref{pds}a-c) the distribution $P(s)$ reveals a strong
deviation from the Poisson. Specifically, one can detect a {\it
clustering} of  energy levels for very small $s$, that results in
a huge peak in the distribution at the origin $s=0$. The presence
of this peak indicates that for weak interaction our model is
highly non-generic and may be compared to integrable 1D-models.
With an increase of the interaction, data for $P(s)$ reveal,
first, a transition to the Poisson, and after, to the
WD-distribution. The more detailed analysis of the data shows that
the transition from  Poisson to  WD-distribution occurs when
the central energy band starts to overlap with the nearest bands.

\section{Theory}

Let us now discuss our model (\ref{ham0}) from the viewpoint of
the standard approach to interacting particles in isolated systems
(see, for example, \cite{var99,I01} and references therein). In
this approach the Hamiltonian is written in the form $H=H_0 + V_0$
where $H_0$ stands for non-interacting particles, and $V_0$
describes a two-body interaction between  particles. The onset of
chaos is usually meant to occur when the strength of the
interaction $V_0$ exceeds the mean energy spacing $d_f$ between
those many-particle states which are {\it directly coupled} by the
interaction. It is important to note that this spacing is much
larger than the mean level spacing $D$ between many-particle
states. Indeed, while  the total density $\rho=D^{-1}$ of states
increases exponentially with the total energy, the density
$\rho_f=d_f^{-1}$  increases only algebraically (for details see,
e.g., \cite{all}).

In order to apply this approach to our model (\ref{ham0}), one
needs, first, to present the Hamiltonian as a sum of the
``unperturbed" Hamiltonian $H_0$, and the part $V=JV_0$ that
stands for the interaction between particles. In our case the
Hamiltonian (\ref{ham0}) can be rewritten in the form,
\begin{equation}
\label{H0V}H= H_0 + JV_0.
\end{equation}
where
\begin{equation}
\label{H00}H_0 =\sum_{k=0}^{L-1} \Big [-\delta_kI^z_k+\Omega I^y_k\Big]
; \,\,\,\,\,\,\, V_0 = -2\sum_{k=0}^{L-2} I^z_k I^z_{k+1}.
\end{equation}

As one can see, the Hamiltonian $H_0$ stands for a kind of {\it
mean field} which absorbs the $\Omega$-dependent term. In this way
the ``mean field " $ H_0$ describes a {\it regular} part of the
total Hamiltonian, and the term $V $, describing the interaction
between the particles, is responsible for chaotic properties ({\it
if any}) of the system. Such a mean-field approach is typical in
the study of chaotic properties of complex atoms and heavy nuclei
\cite{Ce,nuclei}.

\subsection{Delocalization border}

Now, one needs to represent the Hamiltonian (\ref{H0V}) in the
basis in which it is diagonal in the absence of the interaction
($J=0$). In this representation (corresponding to the rotating
basis) the Hamiltonian $H_0$ can be written as a sum of $L$
individual Hamiltonians $H_k$ describing non-interacting {\it
quasi-particles } \cite{qcomp},
\begin{equation}
\label{H0}H_0 = \sum\limits_{k=0}^{L-1} H_k = \sum\limits_{k=0}^{L-1} \sqrt{
\delta_k^2 + \Omega^2 } \,\, I_k^z.
\end{equation}
Correspondingly, in the basis of $H_0$ the interaction $V_0$
between quasi-particles has the form,
$V_0=V_{diag} +
V_{band} + V_{off}$ , where

\begin{equation}
\label{Vvd}
\begin{array}{ll}
V_{diag} = -2 \sum\limits_{k} b_k b_{k+1} I^z_k I^z_{k+1} \,;\,\,\, &
~~~~~~~~ \\
V_{band} = -2 \sum\limits_{k} a_k a_{k+1} I^y_k I^y_{k+1} \,;\,\,\, &
~~~~~~~~ \\
V_{off} = 2 \sum\limits_{k} \left( a_k b_{k+1} I^y_k I^z_{k+1} + a_{k+1}
b_{k} I^z_k I^y_{k+1} \right).
\end{array}
\end{equation}
where
\begin{equation}
\label{ak}b_k = \frac{-\delta_k}{\sqrt{ \delta_k^2 + \Omega^2 }}; \,\,\,\,\,
a_k = \frac{\Omega} {\sqrt{ \delta_k^2 + \Omega^2 }}.
\end{equation}

From Eq.(\ref{H0}) one can see that the energies $\epsilon _k$ of
quasi-particles (or, the same, energies of single-particle states
determined by the Hamiltonian $H_0$) are given by the expression,
\begin{equation}
\label{epstot}\epsilon _k=\pm \frac 12\sqrt{\delta _k^2+\Omega ^2}.
\end{equation}
Note that this relation is valid for any kind of magnetic field
$B^z$ (any dependence $\delta _k$), including the homogeneous
magnetic field ($\delta _k={\it constant}$).

Let us now consider the constant gradient magnetic field
($\delta_k=ak$) for large values of $\Omega \gg \delta_k$. In this
case, one can write an approximate relation for $\epsilon_k$,

\begin{equation}
\label{plusminus}\epsilon _k= \pm \frac 12\left( \Omega \;+\frac{a^2 k^2}{
2\Omega }\right).
\end{equation}
This expression allows one to find global properties of the
unperturbed ($J=0 $) energy spectrum, briefly discussed in the
previous section. Indeed, for large values of $\Omega$ (more
correctly, for $\Omega \gg ak$) one can see that the spectrum has
a band structure, with the bands centered at $0,\,\pm\Omega,\,\pm
2\Omega, ... , \pm L\Omega$.

The central band is defined by such locations of quasi-particles
in the single-particle spectra defined by $\epsilon _k$, for which
an equal number $L/2$ of quasi-particles have positive and
negative values of $\epsilon _k$ (for an even number $L$ of
qubits). Therefore, the total number $N_{cb}$ of many-body states
in the central band is given by the total number of combinations
of $N$ objects having half positive and half negative values,
\begin{equation}
\label{Ncb}N_{cb}=\frac{L!}{(L/2)!(L/2)!}.
\end{equation}
\noindent
One can also see that for $\delta _k=ak$ and $J=0$, the size of the central
energy band is given by twice the maximum energy inside the band,
\begin{equation}
\label{die}(\Delta E)_{cb}=2E_c^{(max)}=2\frac{a^2}{4\Omega }\left[
\sum_{k=L/2}^{L-1}k^2-\sum_{k=0}^{L/2-1}k^2\right]
\end{equation}
$$
=\frac{L^2(L-1)a^2}{8\Omega }
$$

Now, let us discuss the structure of the Hamiltonian matrix
determined by the off-diagonal terms (\ref{Vvd}). One can see that
in the unperturbed basis the term $V_{diag}$ is clearly diagonal.
The action of  $ V_{band}$, is much more complicated. Let us
consider, for simplicity, the central band. Each operator $I_k^y$
flips the k-th spin. Since the interaction is  two-body,
we should consider the action of $ I_k^yI_{k+1}^y$ upon
states as $|...,0_{k+1},1_k,...\rangle $ ,
$|...,1_{k+1},0_k,...\rangle $ , $|...,0_{k+1},0_k,...\rangle $
,$|...,1_{k+1},1_k,...\rangle $. First two kinds of the states
upon the action of $V_{band}$ still remains in the same central
band since the number of $0$'s and $1$'s is conserved. Second pair
of states increases (or decreases respectively) the number of
$1$'s of two units: that means that such a coupling refers to a
next to nearest energy band (nearest bands differ by plus/minus
one $1$'s). As a result, one can conclude that the term $V_{band}$
stands for the interaction both {\it inside} the central band, and
between the {\it next-neighbor} energy bands.

\begin{figure}
\epsfxsize 6cm
\epsfbox{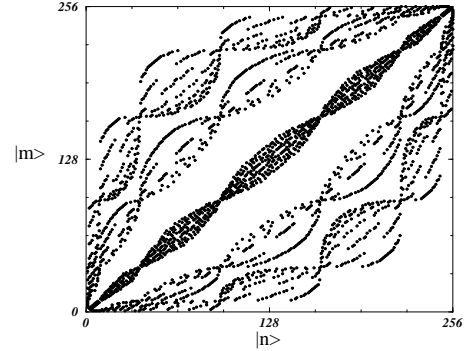}
\narrowtext
\caption{
Structure of the Hamiltonian matrix in the mean field basis for
the $N$-interaction; black points stand for matrix elements whose
modulus is larger that $10^{-6}$. Here is $L=8$,   $\Omega=100$,
\  $J=1$, $\omega_k = \omega_0 + k $. }
\label{matel}
\end{figure}

In the same way it is easy to understand that the term $V_{off}$ give rise
only to the off-band interaction, to be more precise, to a coupling between
{\it nearest} bands. The structure of the Hamiltonian in the mean-field
basis is shown in Fig.\ref{matel}.

For a relatively weak interaction, the eigenstates in the
mean-field basis defined by the unperturbed Hamiltonian $H_0$, are
delta-like functions with an admixture of other components with
small amplitudes. In this case one can speak about the {\it
localization} of eigenstates in the unperturbed basis. With an
increase of the interaction strength, the number $N_{pc}$ of basis
components with large amplitudes ({\it number of principal
components}) increases. According to the theory of interacting
particles, the transition from strongly localized ($N_{pc}\approx
1$) to {\it delocalized} (or, {\it extended}) states (with
$N_{pc}\gg 1$) occurs very fast with an increase of the
interparticle interaction. For this reason, one speaks about the
delocalization transition (in the finite-size basis), see,
e.g.\cite{FI97} and references therein.

Generically, in the models with two-body random interaction $V$
between particles \cite{all}, extended eigenstates with large
$N_{pc}$ turn out to be {\it chaotic}. By this term we mean the
situation when the components of the extended states can be
treated as random and independent quantities. Similar situation
(the onset of {\it quantum chaos}) occurs in many dynamical
systems with complex enough interaction, such as many-electron
atoms and heavy nuclei \cite{Ce,nuclei}. In  these systems, the
delocalization transition coincides with the transition to chaos,
and is determined by the condition $V\geq d_f$ (with $V$ as a
typical interaction strength, and $d_f$ as the mean energy
distance between directly coupled many-body states).

Let us now discuss  the delocalization transition in our model,
keeping in mind that it can be different from the transition to
chaos. As it will be shown, our model with the $N$-interaction
manifests a quite unexpected phenomena, namely, the above two
transitions turn out to be very different.

We start with the estimate of the mean level spacing $d_f$ in the
central energy band of our model (\ref{H0V}), between the
many-body states coupled by the interaction (\ref{Vvd}). The
energy spacing $d_f$ can be estimated as the ratio
\begin{equation}
\label{deltaEf}d_f \approx \frac {(\Delta E)_f}{M_f}
\end{equation}
where $M_f$ is the number of many-body states coupled by
$V_{band}$ inside the energy interval $(\Delta E)_f$. In fact,
$M_f$ is the mean number per line of non-zero off-diagonal
elements in the total Hamiltonian (\ref{H0V}).

In order to estimate $M_f$, we note that the interaction
$V_{band}$ in the central band can only couple  those many-body
states having an equal number ($L/2$) of spins ``up'' and ``down''
(for an even number $L$ of qubits). The minimal value of $ M_f=1$
corresponds to the state $$ |0_{L-1},...,0,1,...,1_0\rangle , $$
and the maximal one, $M_f=L-1$, corresponds to the state $$
|0_{L-1},1_{L-2},0_{L-3},1_{L-4},...,0_1,1_0\rangle . $$ Indeed,
in the first case there is only one possibility to change $0$ to
$1$ , and $1$ to $0$ for the {\it nearest} qubits. And in the
second case, there are $L-1$ such changes, each of them
corresponding  to the nearest neighbor interaction with no  change
in the total number of spins ``up''  and ``down''. Therefore, one
can estimate the average value $M_f$ as $ M_f\approx L/2$ which is
in a very good agreement with direct numerical check.

\begin{figure}
\epsfxsize 6cm
\epsfbox{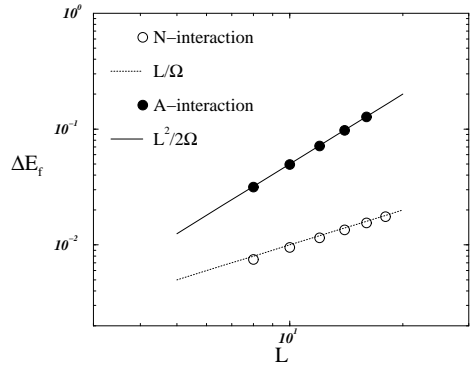}
\narrowtext
\caption{
Numerical calculation of $(\Delta E)_f$ for both  $N$- (open
circles) and $A$-interaction (full circles), see Sect.VI) showing
the $L$-dependence as found analytically. Here is $a=1$. }
\label{d}
\end{figure}

One should stress that the energy range $(\Delta E)_f$ within
which the many-body states are coupled, is much less than the
total energy width $ (\Delta E)_{cb}$ of the central band
determined by Eq.(\ref{die}). The value of $(\Delta E)_f$ can be
estimated as the maximal difference between energies
$E_c^{(2)}=\sum_k^{(2)}\epsilon_k $ and $E_c^{(1)}=\sum_k^{(1)}
\epsilon_k$ of two many-body states $|\psi_1\rangle$ and $|\psi_2\rangle$
of $H_0$, having the matrix element  $\langle
\psi_1|V_{band}|\psi_2\rangle$ different from zero. If we consider
only the coupling inside the central band we can find these values
$E_c^{(2)}$ and $E_c^{(1)}$ by observing that the maximal energy
is obtained by flipping the outermost spins. Application of
$I_{L-1}^y I_{L-2}^y$ to the state $$ |\psi_1 \rangle
=|1_{L-1},0_{L-2},...\rangle $$ gives rise to the state $$ |\psi_2
\rangle =|0_{L-1},1_{L-2},...\rangle; $$ ($L-1$ and $L-2$
correspond to the states with the highest values of
single-particle energies $\epsilon_k$). Thus, the energy
difference $ E_{|\psi_1 \rangle} - E_{|\psi_2 \rangle}$ is given
by $$ (\Delta E)_f = \frac{a^2}{4\Omega}2 [(L-1)^2 - (L-2)^2] =
\frac{ a^2}{\Omega} (L-\frac{3}{2}) $$ Numerical results confirm
this prediction very well, see Fig.\ref{d}.

As a result, for $L \gg 1$ we have,
\begin{equation}
\label{deltae}d_f = \frac {(\Delta E)_f}{M_f}
\approx \frac{2a^2}{\Omega
}
\end{equation}

The mean spacing $d_f$ should be now compared with the typical
value of the perturbation, $V=JV_0$. The latter can be found from
$V_{band}$ as $V\approx J/2$ (other terms are negligible  for
$\Omega \gg \delta _k$). Therefore, we finally obtain,
\begin{equation}
\label{J1}J_{cr}\approx \frac{4a^2}\Omega .
\end{equation}

Surprisingly, the delocalization border does not depend on the
number of qubits, in contrast to the result of \cite{dima} where
$J_{cr}$ decreases as $1/L$. The reason is the  specific influence
of a constant gradient of the magnetic field that results in the
quadratic dependence $\epsilon_k \propto k^2$ for the
single-particle levels of quasi-particles of the mean-field
Hamiltonian $H_0$ (see Eq.(\ref{H0})).

Let us now compare the analytical estimate (\ref{J1}) with numerical data.
The commonly used quantity to measure the number $N_{pc}$ of principal
components in eigenstates, is  the so-called {\it inverse
participation ratio},
\begin{equation}
\label{ipre}N_{pc}(E) = \left[\sum_n |\psi_n(E)|^4\right ]^{-1}.
\end{equation}
Here $\psi_n(E) = \langle n | \psi(E)\rangle$ is the $n-th$ component of a
particular eigenfunction corresponding to the eigenvalue $E$.

From Eq.(\ref{ipre}) one can see that for equal values of the
components of an eigenstate, $\psi _n=1/\sqrt{N}$, the number of
principal components is equal to the size of the basis,
$N_{pc}=N$. In another extreme limit of completely extended and
chaotic eigenstates, the value of $N_{pc}$ is equal to $N/3$. The
factor $3$ arises due to the Gaussian fluctuations of $\psi _n$
which are generic in the case of strong quantum chaos (see,
e.g.\cite{GMW99} ). For localized states the value of $N_{pc}$
approximately gives the number of basis states effectively
occupied by this eigenstate.

\begin{figure}
\epsfxsize 7cm
\epsfbox{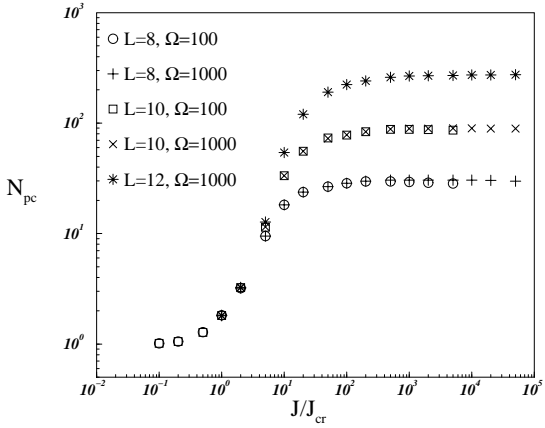}
\narrowtext
\caption{
The average number of principal components
in the rotated basis for the eigenstates from the
central energy band, as a function of $J/J_{cr}$ for $L=8,\,10,\,12$ and
different values of $\Omega$. }
\label{dyn}
\end{figure}

Numerical data for $N_{pc}$ computed in the mean field basis
(where $H_0$ is diagonal for $J=0$, see Eq.(\ref{H0V})) for the
eigenstates taken from the central energy band, are given in
Fig.\ref{dyn} in dependence on $J/J_{cr}$. It is clearly seen that
below the delocalization border, $J < J_{cr}$, there is a scaling
dependence of $N_{pc}$ on $L$ and $\Omega$ in accordance with the
estimate (\ref{J1}). On the other side, for $J \gg J_{cr}$, the
value of $N_{pc}$ saturates to its maximal value $N_{cb}/3$ in
correspondence with random matrix predictions (here $N_{cb}$ is
the total number of states inside the central energy band, see
(\ref{Ncb})). The latter correspondence of the maximal value of
$N_{pc}$ to $N_{cb}/3$ is a strong evidence of the quantum chaos
in the model for a very large interaction.

\subsection{Chaos border}

In this section we study the transition to global chaos which is
due to the overlapping of the energy bands. In order to obtain the
condition for the band overlapping, one needs to find the band
width and to compare it with the distance $\Omega$ between the
bands.

We have shown that in the absence of the interaction, the energy
width of the central band can be estimated analytically, see
Eq.(\ref{die}). Numerical data reported in Fig.\ref{bband}, show
that with an increase of the interaction, the band width $\Delta
E$ saturates to some value $(\Delta E)_s$ which is independent of
$\Omega $. Therefore, we can estimate $(\Delta E)_s$ by coming
back to the $z-$representation of Hamiltonian (\ref{ham0} )
where the $\Omega $-depending term enters in the off-diagonal
matrix elements only. By omitting this term, we can write the
diagonal part,
\begin{equation}
\label{hamd}H_d=-\sum_{k=0}^{L-1}\left[
akI_k^z+2J\sum_{k=0}^{L-2}I_k^zI_{k+1}^z\right] .
\end{equation}
where the relations $\omega _k=\omega _0+ak$, and $\nu =\omega _0$
are directly taken into account.

In the case of $a L \gg J$ we are interested in, the unperturbed
($J=0$) many-body energy spectrum of (\ref{hamd}) is given by a
sequence of degenerate levels separated by the spacing $a$. Due to
a weak interaction $J$ , each set of these degenerate levels
spreads and creates the energy bands. In order to find  the
central energy bandwidth, we should consider the action of the
interaction operator $$-2J\sum_{k=0}^{L-2} I^z_k I^z_{k+1}$$ upon
the states belonging to the central band The latter assumption is
an approximation: in  $z$-representation, the central energy band
can also contain few states with slightly different numbers of 0's
and 1's.

The action of each term in the interaction operator leaves the
state as it is, multiplying it by a factor $\pm J/2$ depending on
the presence of two close $ 11$ and $00$ or different $01$ and
$10$. This results in the shift of the energy from its zero-value
in the central band. Two configurations, $ |-\rangle
=|0,0,...,0,1,1,...,1\rangle $ and $|+\rangle
=|0,1,0,1,...,0,1\rangle $, should be considered which correspond
to the maximal shift in the ``negative'' and ``positive''
directions. In this way we can safely say that such a bandwidth is
given by the energy difference $ E_{|+\rangle }-E_{|-\rangle }$.
It is easy to see that one has $$ E_{|+\rangle }=\frac{(L-1)Ja}2
$$ and $$ E_{|-\rangle }=\frac{-(L-3)Ja}2 $$ thus giving
\begin{equation}
\label{dej}\Delta E=(L-2)Ja\qquad {\rm for}\qquad \Omega \gg J
\end{equation}
By equating two expressions Eq.(\ref{die}) and (\ref{dej}) we finds the
transition point
\begin{equation}
\label{Jb}J_0\approx \frac{L^2a}{8\Omega }
\end{equation}
between the two dependencies for the band width $\Delta E$.

One can see that for $J>J_0$  bands are overlapped if $(L-2)Ja\geq
\Omega $. That gives the critical value $J_b$ for the overlapping,
\begin{equation}
\label{Jo1}J_b\approx \frac \Omega {aL}.
\end{equation}
subject to the condition $J>J_0$. By comparing Eqs.(\ref{Jb}) and
(\ref{Jo1}), one gets the lower bound $J\geq
\sqrt{L/8}$ compatible with the above two constraints.

On the other side, one can also have the band overlapping whenever
$J<J_0$, if $a^2L^2(L-1)/8\Omega \geq\Omega$. Therefore, in this
case the overlapping of the bands occurs for any $J$, if the
number of qubits is large enough, $ L\geq 2(\Omega/a)^{2/3}$.

One should stress that overlapping of bands is not a sufficient
condition in order to have the delocalization of eigenstates.
Indeed, the estimate Eq.(\ref{J1}) for the delocalization border
$J_{cr}$ is derived for the central energy band only, therefore,
it is not valid when bands are overlapped. Therefore, one needs to
start with the expression (\ref{deltaEf}) and estimate $(\Delta
E)_f$ for the case when the energy spectrum is not a band-like.

In order to do this, it is convenient to switch to the mean-field
representation with the unperturbed Hamiltonian $H_0$ given by
Eq.(\ref{H00} ). The total size of the unperturbed energy spectrum
is now defined by the difference between the energies
corresponding to the following limiting configurations, $$
|\uparrow \rangle = |1_{L-1},1_{L-2},...,1_1,1_0 \rangle $$ and $$
| \downarrow \rangle = |0_{L-1},0_{L-2},..., 0_1,0_0 \rangle. $$
But this is not what we need. Indeed, these two many-body states
are not coupled by the {\it two-body} interaction (\ref{Vvd}).
What we need to find, is the maximal energy change due to the
action of the $J$-interaction. To do that we have to consider two
states corresponding to the flipping of both the two uppermost
spins, namely  : $$ |\uparrow\rangle =  |1_{L-1},1_{L-2},...
\rangle $$ and $$
| \downarrow \rangle =  |0_{L-1},0_{L-2},... \rangle. $$

\noindent
The energy difference between such states is given by :
$$
(\Delta E)_f \approx 4\sqrt{ a^2 L^2 +\Omega^2}.
$$
Since the number of coupled states remains the same, $M_f\simeq L/2$, one
gets that in order to have the transition to delocalized states in the case
of the overlapped bands, the typical value of the interaction has to be
larger than $d_f = (\Delta E)_f/M_f$,
$$
J/2 \geq \frac{ 4 \sqrt{ a^2 L^2 + \Omega^2}}{L/2}
$$
or
\begin{equation}
\label{geic}J \geq J_c \simeq \frac {16}{L} \sqrt{ a^2 L^2 + \Omega^2}
\end{equation}

One should notice that the two criteria (band overlapping and transition to
delocalization), if both satisfied, are expected to result in the onset of
chaos. This conclusion is confirmed numerically, and is supported by
analytical arguments.

Indeed using data   from Fig.\ref{pds}, e.g $L=12$, $\Omega=100$,
$a=1$, one gets a chaos border $J_c \approx 130 $ well confirmed
by the Wigner-Dyson distribution in Fig.\ref{pds}f.

On the other hand, we have already seen that the simple
requirement to be in the delocalized regime $(J>J_{cr})$ without
the overlapping of bands, does not give rise to chaos in our
dynamical model with the nearest interaction.

It is also easy to check that the conditions of the band
overlapping for the case $J< J_0 = \Omega / 8ax$ and $L> 8 x$
(with $x=\Omega^2 /L^2 a^2$ ), are not compatible with the
delocalization border $J > J_c = 16 a \sqrt{1+x}$ in the region of
non-selective excitation, $x \gg 1 $. This means that a relatively
weak interaction does not lead to the delocalization (and,
therefore, to the chaos), in spite of the overlapping of the
energy bands.


\section{Structure of eigenstates in the $z$-representation}

The analytical treatment we have performed in the previous
section, is based on the mean-field representation of our model,
namely, when the Hamiltonian matrix is written in the basis of the
``unperturbed" part $H_0$, see Eq.(\ref {H0}). This approach is
natural for the theoretical study since the interaction is much
less that the $\Omega$-dependent term ($J \ll \Omega$), therefore,
the interaction between qubits can be considered as a weak
perturbation.

However, the dynamical properties of the model are related to the
$z$-representation which is adequate to the experimental setup.
For this reason  we discuss below the structure of eigenstates of
Hamiltonian (\ref{ham0} ) in  $z$-representation, in
relation with the above analytical estimates obtained in the
mean-field approach.

\begin{figure}
\epsfxsize 7cm
\epsfbox{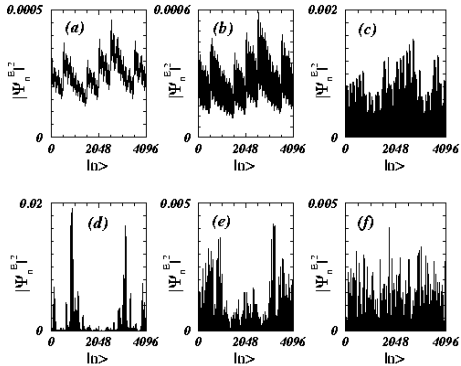}
\narrowtext
\caption{
Typical structure of eigenfunctions for different interaction strengths,
$J=0,0.0002,0.1,1,10,100$ denoted by (a,b,c,d,e,f) respectively. Eigenstates
are taken from the central energy band for $L=12,\Omega =100$ , $\omega
_k=\omega _0+k$, $k=1,...,L$, $\omega _0=100$, $\nu =\omega _0$, $a=1$. }
\label{eig}
\end{figure}

Since the most important question is about the role of the
interqubit interaction,  main attention is paid to the
dependence of global properties of eigenstates on the interaction
strength $J$. Typical structure of the eigenstates in the
$z$-representation is shown in Fig.\ref{eig} for different values
of $J$. First, one should note that in this basis all components
of eigenstates in the absence of the interaction, $J=0$, are very
close, in average, to $|\psi_n|=1/\sqrt N$. If the interaction is
very weak, the standard perturbation theory is valid, and a kind
of fluctuations of the probabilities $w_n=|\psi_n|^2$ is expected
around the mean value $w_n=1/N$ where $N$ is the total size of the
basis (the total number of many-particle states).

The data show that if the interaction $J$ is relatively strong,
the components of eigenstates are quite different from the
unperturbed values. This region may be very important for 
quantum computation, and the main problem is to know whether these
errors in the components of the eigenfunctions (the deviations
$\delta w_n$ from the unperturbed value $1/N$) can destroy quantum
coherent effects needed for the quantum computation. This problem
was addressed in our previous study \cite{gena}, here we are
mainly interested in global properties of eigenstates for a very
broad region of the interaction.

The most interesting conclusion which can be drawn from the
numerical data for a weak enough interaction (see
Fig.\ref{eig}(a-b)), is that the eigenstates turn out to have a
regular structure, even if the deviations $\delta w_n$ are
relatively large. Indeed, one can see regular global dependence of
$w_n$ on the basis number $n$, with some fluctuations around the
mean. This fact seems to be directly related to the specific
structure of the Hamiltonian matrix presented in Fig.\ref{mat}.

With an increase of the interaction, the regular structure of
eigenstates disappears and huge fluctuations of components of
eigenstates emerge, see Fig.\ref{eig}(c-e). The structure of these
eigenstates is very similar to that known in the physics of
disordered systems, when eigenstates ``occupy'' some fraction of
the basis, without noticeable correlations between different
components $\psi _n$ (see for example, \cite{I01} and references
therein). One can say that these eigenstates are {\it sparse} in
the sense that the number $N_{pc}$ of principal components of the
eigenstates is much less than the total size $N$ of the basis.
Therefore, there is a strong change in the structure of
eigenstates (compare Figs.\ref{eig}(a-b) with
Figs.\ref{eig}(c-e)). One can say that the transition from extended
{\it regular states} to the {\it weakly chaotic states} occurs for
$J\approx 0.1$.

When the interaction between qubits increases further, one can see
another transition to {\it strongly chaotic states}, see
Fig.\ref{eig}(f). The latter are characterized by an {\it ergodic}
filling of the whole basis, and by strong fluctuations of the
components $\psi _n$, which are found to be practically random and
independent. This situation is well described by RMT
(see, e.g.\cite{I90}). Therefore, for such strong
interaction $J\approx 100$, chaotic properties of our system are
very strong and quantum computation process can be destroyed.

In order to quantitatively characterize the eigenstates, we have computed
the number $N_{pc}$ of principal components defined by Eq.(\ref{ipre}).
Another measure of the spread of an eigenstate in a given basis, is its
''width'' $\sigma (E)$ determined as
\begin{equation}
\label{wid}\sigma (E)=\left[ \sum_n|\psi _n(E)|^2n^2-\left( \sum_n\ n|\psi
_n(E)|^2\right) ^2\right] ^{1/2}
\end{equation}
Note, that in contrast to $N_{pc}$ which gives an effective number
of large components, and is insensitive to the location of these
components, the width $\sigma (E)$ does not ''feel'' the presence
of ''holes'' in the sparse eigenstates. The latter fact can be
used to distinguish chaotic {\it ergodic} states from the sparse
ones. Namely, for fully extended but very sparse eigenstates, the
value of $\sigma (E)$ is of the order of $N$, however, $N_{pc}$ is
much less than $N$.

\begin{figure}
\epsfxsize 6cm
\epsfbox{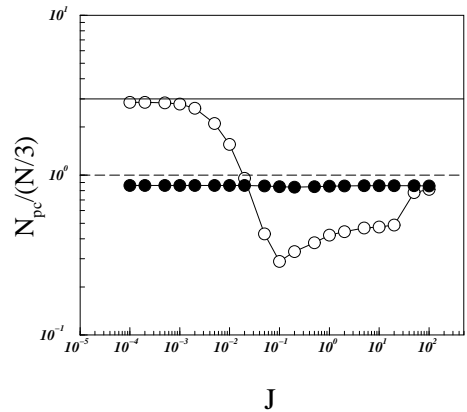}
\narrowtext
\caption{
Normalized average number of principal components $N_{pc}$ (open
circles) and the width $\sigma$ (full circles) as a function of
$J$ in $z$-representation for $\Omega=100$. 
The average is taken over the
eigenfunctions from the central band only. The solid horizontal
line corresponds to $N_{pc} = N$, and the dotted line gives the
extreme limit of completely chaotic and extended states,
$N_{pc}=N/3$. The parameters are the same as in Fig.\ref{eig}. }
\label{lln}
\end{figure}

The mean values of $N_{pc}$ and $\sigma$  in
dependence on the interaction $J$ are given in Fig.\ref{lln}. The
circles represent the value of $N_{pc}$ and $\sigma$, averaged
over the eigenstates from the central energy band. First of all,
one should note that the width $\sigma$ turns out to be large and
independent of the interaction. This means that all eigenstates
are {\it extended} in the $z$-representation, in spite of a
serious difference in their structure, see Fig.\ref{eig}.
Contrary, the number of principal components $N_{pc}$ demonstrates
two principal transitions in the structure of eigenstates.

Numerical data of Figs.\ref{eig}-\ref{lln} allows one to
distinguish between few different regions of the interaction
strength $J$. The first region with a very weak interaction $J\leq
2\,\cdot 10^{-3}$ is characterized by the constant value
$N_{pc}\approx N$ and corresponds to completely extended ($ |\psi
_n|^2\approx 1/N$) eigenstates shown in Fig.\ref{lln}(a-b). In
this region the energy spectrum consists of many close
quasi-degenerate levels, thus leading to a strong deviation from
the Poisson distribution, see Sect.III.

\begin{figure}
\epsfxsize 7cm
\epsfbox{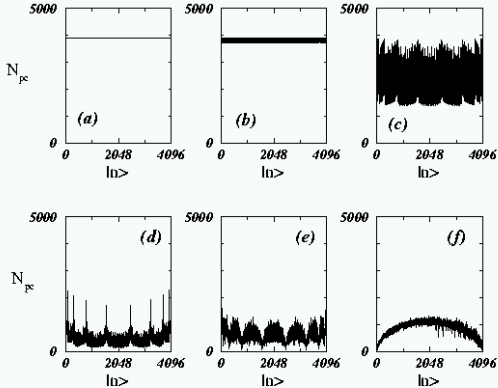}
\narrowtext
\caption{
Number of principal components $N_{pc}$ for all eigenstates
reordered in increasing energy ($|0\rangle$ is the ground state,
$|1\rangle$ is the first excited state, etc.). Data correspond to
the parameters of Fig.\ref{eig}. }
\label{ipr}
\end{figure}

In the second region with $N_{pc} \ll N$, all eigenstates are
strongly influenced by the inter-qubit interaction. This region
was termed in Ref.\cite{qcomp} the region of {\it weak chaos}
since the structure of eigenstates looks chaotic (see
Fig.\ref{eig}d), however, the level spacing distribution $P(s)$ is
quite close to the Poisson. From the data, the transition to the
weak chaos occurs for $J \approx 0.05$ and corresponds to the
analytical estimate (\ref{J1}) for the delocalization transition
in the mean-field basis. The very point is that the critical value
$J_{cr}$ given by Eq.(\ref{J1}), in the $z$-representation
corresponds to the transition from completely extended states to
the {\it weakly chaotic} states. One should stress that from the
practical point of view the region of weak chaos may be dangerous
for quantum computation because of large deviations of eigenstates
from the unperturbed ones, see Fig.\ref{eig}(c-d).

Second transition to  {\it strong quantum chaos} occurs for $J
\sim 100$. By the latter term we denote the situation when the
level spacing distribution has the Wigner-Dyson form and
fluctuations of components $\psi_n$ are close to Gaussian ones
with $N_{pc} \approx N/3$, see Fig.\ref{lln}. As we already
discussed, this transition corresponds to the simultaneous
occurrence of both band overlapping and delocalized
states, see  (\ref{geic}). One can see that strong quantum
chaos for  $N$-interaction emerges for an extremely strong
interaction and thus it is not relevant 
for  quantum computation.

More detailed information about the global structure of
eigenstates can be drawn from Fig.\ref{ipr} where the value of
$N_{pc}$ is shown for all eigenstates $\psi_n(E^{(m)})$ reordered
in increasing energy $E^{(m)}$. In this figure one can see how the
band structure of the spectrum manifests itself in the value of
$N_{pc}$. In particular, it is seen that for non-overlapped bands
there is a quite strong dependence of $N_{pc}$ on whether the
energy $E^{(m)}$ of a specific eigenstate is at the center of
energy bands or close to the band edges.

One should point out a remarkable difference for the behavior of
$N_{pc}$ close to the band edges,  compare Fig.\ref{ipr}d and
Fig.\ref{ipr}e. Namely, in the region of parameters of
Fig.\ref{ipr}d, the highest value of $ N_{pc}$ corresponds to the
band edges, in contrast to Fig.\ref{ipr}e where at the band edges
the eigenstates are extremely localized (with a very small value
of $N_{pc}$). The origin of this difference is not clear, however,
it should be noted that the data reported in Fig.\ref{ipr}e have
already been observed (and explained) in few  models of isolated
systems with interacting particles (see, for example, \cite{faus}
and \cite{german}). For those models it was found that for the
unperturbed eigenstates which are close to the band edges, the
interaction with other basis states is strongly suppressed.

\section{Random Models}

In the previous Sections we have discussed the {\it dynamical}
model (\ref {ham0}) of interacting qubits. We have seen that in
spite of the absence of any randomness in this model, for a very
strong interaction both energy spectra and structure of
eigenstates reveal chaotic properties which are generic for
quantum chaos. In this sense, it is interesting to compare the
obtained results with those for similar models with random
interaction. This problem is not academic since in reality there
are many effects which can lead to some  randomness in the
Hamiltonian (\ref{ham}).

\subsection{All-to-all interaction}

It is instructive to see what happens for a long-range interaction between
qubits. We have studied in details the case when the  interaction
couples all qubits in the same manner ($A$-interaction),
\begin{equation}
\label{Nint}H=\sum_{k=0}^{L-1} \left[-\delta_k I^z_k+ \Omega I^y_k -2 \
\sum\limits_{n>k}^{}J_{k,n}I^z_k I^z_n \right].
\end{equation}
Here the interaction is assumed to be completely random, with
$J_{k,n}=J\xi$ where $\xi$ are random numbers with a flat
distribution inside the interval $[-1,+1]$.

This model can be treated  analytically in the same way as we did
it in Sect.IV. Specifically, we are interested in the
delocalization border which is determined by the comparison of the
ratio (\ref{deltaEf}) with the typical interaction strength.

\begin{figure}
\epsfxsize 8cm
\epsfbox{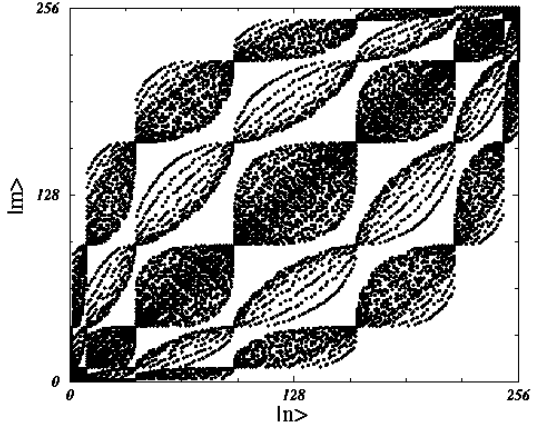}
\narrowtext
\caption{
Structure of the Hamiltonian matrix in the mean field basis for
the $A$-interaction; black points stand for matrix elements whose
modulus is larger that $10^{-6}$. Here is $L=8$,   $\Omega=100$,
\  $J=1$, $\omega_k = \omega_0 + k $. }
\label{mataa}
\end{figure}

The modification of the Hamiltonian (\ref{H0V}) written in the
mean field basis is straightforward. Specifically, the structure
of the unperturbed part, see Eq.(\ref{H0}), remains the same, and
the interaction term (\ref{Vvd}) has the same structure (the only
difference being the summation taken over all qubits). The most
important point is  that the Hamiltonian matrix has a different
structure from that for the $N-$interaction, see Fig.\ref{mataa}

Despite the block structure shared by the analogous matrix for the
$N$-interaction, shown in Fig.(\ref{matel}), and due to  two-body
interaction, each block is now characterized by many elements
different from zero. For this reason, one can expect that chaotic
properties of the model with the $A$-interaction are much stronger
that those found in the case of $N$-interaction.

The estimate for $M_f$ can be obtained for the $A$-interaction as well.
Since all qubits are allowed to interact to each other, the maximum number
of coupling between unperturbed many-body states inside the central energy
band with all the others, is
\begin{equation}
\label{Amf}M_f = \frac{L^2}{4}.
\end{equation}

As for  $(\Delta E)_f$, it can be found by
considering the maximal energy shift obtained by applying
the operator $I_0^y I_{L-1}^y$ to the state $
|1_{L-1},...,0_0\rangle,
$ and resulting in the new state $ |0_{L-1},...,1_0\rangle. $ The
energy difference between these two states is given by $$ (\Delta
E)_f \simeq \frac{a^2}{4\Omega} (2 L^2 ), $$ which perfectly
agrees with the direct computations, see Fig.\ref{d}. As a result,
the critical value $J^{a}_{cr}$ for the delocalization border is
obtained from the relation,

$$
\frac{J_{cr}^{a}}{2} \approx \frac{(\Delta E)_f}{M_f} =
\frac{2 a^2}{\Omega},
$$

therefore,
\begin{equation}
\label{J2}J_{cr}^{a} \approx \frac{4 a^2}{\Omega}.
\end{equation}

\begin{figure}
\epsfxsize 8cm
\epsfbox{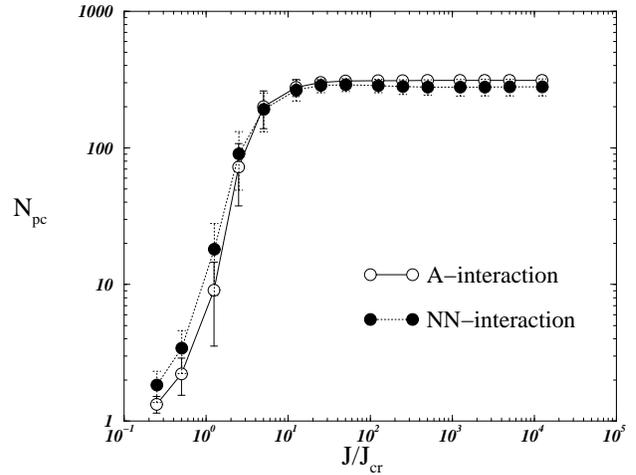}
\narrowtext
\caption{
The average number of principal components in the mean field
basis, for the eigenstates from the central energy band, as a
function of $J/J_{cr}$ for $L=12$ and $\Omega=1000$. Open circles
are for the $A$-interaction, full circles are for the
$NN$-interaction, see next Section. }
\label{ntnaa}
\end{figure}    

This is an  unexpected result since it coincides with the
estimate (\ref {J1}) for the delocalization border in the case of
$N$-interaction. The reason is that the energy range $(\Delta
E)_f$ within which many-body states are connected by the
interaction, and the number $M_f$ of the states within this energy
range, are both proportional to $L^2$. The result shows that the
delocalization border turns out to be independent on the range of
the interqubit interaction.

However, chaotic properties of this random model with the
$A$-interaction are much stronger than those found for the
$N$-interaction. Namely, the chaos border for the $N$-interaction
turns out to coincide with the delocalization border. The
transition to delocalized states for the $A-$interaction is shown
in Fig.\ref{ntnaa}.

\begin{figure}
\epsfxsize 7cm
\epsfbox{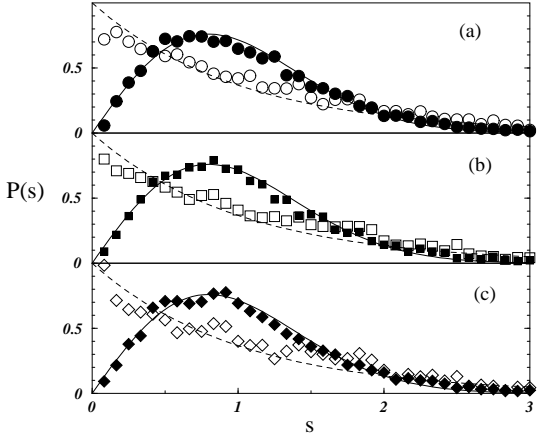}
\narrowtext
\caption{
Level spacing distribution for eigenvalues in the central band for
$L=10$ and $\delta_k=k$. For the average, $30$ different matrices
with the random $A$-interaction have been used. a) $\Omega=10$,
$J=0.1$ (open circles), $J=1$ (full circles); b) $\Omega=100$,
$J=0.01$ (open squares), $J=0.1$ (full squares); c) $\Omega=1000$,
$J=0.001$ (open diamond), $J=0.01$ (full diamond). Note that the
theory predicts a transition point at $J=J_{cr}\sim 4/\Omega$. For
comparison, both the Poisson (dashed line) and the Wigner-Dyson
(full line) distributions are shown. }
\label{ata}
\end{figure}

The closeness of the delocalization and chaos borders for the
$A$-interaction can be also checked by studying the level spacing
distribution. The latter is expected to manifest a transition from
the Poisson to the Wigner-Dyson at the critical value of $J$ given
by the above estimate (\ref{J2}). In Fig.\ref{ata} we show that
the transition to chaos is independent from the product
$J\Omega$, in correspondence with the analytical prediction
(\ref{J2}). These results prove that for the $A$-interaction our
model is similar to generic models for which the delocalization
border coincides with the chaos border.

\subsection{Next to nearest interaction}

Finally, we discuss the intermediate case when the interaction $V$
in the dynamical model (\ref{ham}) couples four next nearest
qubits, $k\pm 1, k \pm 2$, (the $NN$-interaction).

\begin{figure}
\epsfxsize 6cm
\epsfbox{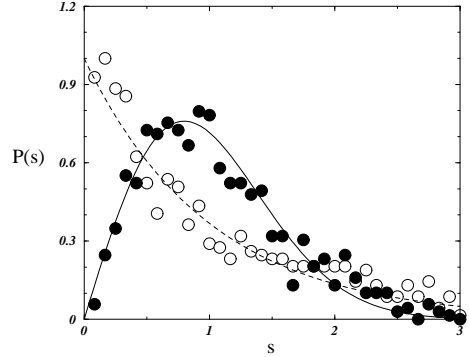}
\narrowtext
\caption{
Nearest neighbor distribution for eigenvalues in the central band
for $L=12$. One single matrix with  random $NN$-interaction have
been used. Open circles are for $J=0.001$, close circles for
$J=1$. For comparison, both the Poisson (dashed line)  and
Wigner-Dyson (full  line) distributions have been  shown. }
\label{pos}
\end{figure}

A straightforward analysis similar to that shown in the previous
Sections leads to the same critical border for delocalized states,
as those found for the $N$ and $A$ interactions. This has been
numerically confirmed, see that data in Fig.\ref{ntnaa}. Moreover,
as for the $A$-interaction, the delocalization border for the
$NN$-interaction turns out to coincide with the chaos border. This
has been proved by using the level spacing distribution, see
Fig.\ref{pos}.

Our numerical study shows that, in contrast to the case of the
$N$-interaction (when only two neighbor qubits are coupled), the
quantum chaos emerges for much lower values of the
$NN$-interaction, for $0.1 < J < 1.0$, see Figs.\ref{pos}. This
region of parameters $J$ and $\Omega$ is important from the
experimental viewpoint, therefore, the quantum chaos may have a
real influence for  quantum computation.

Since any other of long range interactions can be seen within
these two extreme cases (the $A$ and $NN$ interactions), one can
conclude that for a typical interaction (other than strictly
between nearest qubits), the quantum chaos can emerge for a quite
weak interaction and may have an influence on a quantum computer
operability. Therefore, it may be important to reduce the range of
the interqubit interaction in an experimental setup of a quantum
computer.

\section{General discussion}

\subsection{Quasi-integrability}

As we noted, the model (\ref{ham0}) with the interaction between
nearest qubits has quite specific properties. Namely, the
delocalization border turns out to be very different from the
border of quantum chaos. Below we explain this phenomena in terms
of quasi-integrability of our model.

Let us come back to the expression for the off-diagonal matrix
elements of the Hamiltonian (\ref{H0V}) in the mean-field basis
determined by the eigenstates of $H_0$, see Eqs.(\ref{Vvd}). For
the case of large $\Omega \gg\delta _k$, we are interested in, the
term $V_{off}$ is small compared to $V_{band}$ since $a_k \sim 1$
and $b_k\sim  -1/\Omega $. Also, the diagonal term $V_{diag}$ is
much smaller than the two other terms (it is proportional to
$b_k^2\sim 1/\Omega ^2\ll 1/\Omega \ll 1$, see Eq.(\ref {ak})).
Therefore, the approximate Hamiltonian $H_a$ can be written in the
following form
\begin{equation}
\label{Ha}H_a=\sum_{k=0}^{L-1}\gamma
_kI_k^z-\sum_{k=0}^{L-2}J_kI_k^yI_{k+1}^y.
\end{equation}
where $\gamma _k=\sqrt{\delta _k^2+\Omega ^2}$ and $J_k=2J$ for
our model.

This Hamiltonian has been recently studied in a number of papers,
(see, for example, \cite{young} and references therein). It was
shown \cite{lieb} that for independent random variables $\gamma
_k$ and $\xi _k$ the model (\ref{Ha}) can be mapped to an
Hamiltonian describing $L$ free fermions. This transformation
holds only in the case of nearest neighbor coupling. Therefore,
this model is integrable and the level spacing distribution $P(s)$
can be expected to be Poisson--like  for {\it any} interaction
strength $<J_k^2>^{1/2}$. This explains why for non-overlapping
bands our original Hamiltonian (\ref{H00}) with $\Omega \gg \delta
_k$ reveals the Poisson for $P(s)$ above the delocalization
border.

It should be noted that the delocalization border $J_{cr}$ (see
Sect.IV) results from the standard perturbation theory which takes
into account a two-body nature of interaction. Namely, when the
typical interaction which connects unperturbed many-body states is
much larger than the mean distance between energy levels of {\it
these} states, in the corresponding basis the interaction creates
exact eigenstates with many components. Typically, these compound
states are chaotic due to a complex structure of the interaction.
This is why the delocalization border generically coincides with
the quantum chaos border. However, in specific cases like our
quasi-integrable model (for $\Omega \gg \delta _k$ and not very
strong interaction), the delocalization border and the onset of
chaos may be very different.

The above analysis is also helpful in the explanation of the strong
difference between the model with $N$-interaction, from the model
when qubits are coupled by a different kind of interaction ( $A-$
or $NN-$ interaction, see previous Sections). Indeed, in the
latter cases the interaction $V$ has many additional terms
compared with Eq.(\ref{Vvd}), and results in a strong coupling
between all energy bands. This leads to  quasi-integrability
breaking, and to the onset of chaos at the border of
delocalization.

\subsection{Role of magnetic field}

Our approach based on the mean-field representation, see Sect.IV,
is valid for any kind of the $B^z$-magnetic field. Let us consider
the simplest case of a homogeneous magnetic field for which all
frequencies of spin's precession $\omega _k$ are the same, $\delta
\omega _k=\omega _0-\nu =f$. For a {\it non-resonant} case with
$f\neq 0$, and in absence of the interaction ($J=0$), the energy
spectrum has no more a band structure since each of the $L+1$
levels is degenerate. Indeed, each single-particle energy has two
values $\epsilon _k=\pm \frac 12(\Omega +\frac{f^2}{2\Omega })$
only, where $f \ll \Omega$. Since all many-body states in the
central band have the same number of pluses and minuses in the
expression for the total energy, the latter is zero. Thus, the
level spacing $ (\Delta E)_f$ is also zero which means that any
small interaction gives rise to delocalized states.

\begin{figure}
\epsfxsize 7cm
\epsfbox{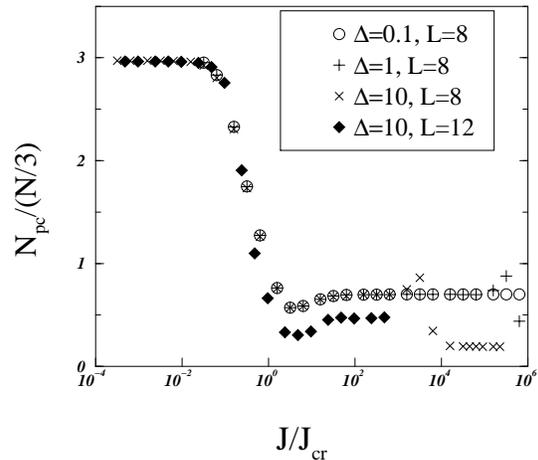}
\narrowtext
\caption{Average number of principal components for eigenfunctions
in the central band for homogeneous magnetic field and random
frequencies in the interval $(\nu-\Delta/2, \nu+\Delta/2)$, versus
the rescaled interaction $J/J_{cr}$, where $J_{cr}$ is defined by
Eq.(\ref{crho}).  }
\label{homo}
\end{figure}

In recent studies \cite{dima} random variation of spin frequencies
is included in the model, in order to take into account effects of
finite temperature and environment. For this reason the energies
are not exactly degenerate but swap into finite width bands. In
the same way, let us assume that the energy of many-body states
fluctuates, thus resulting in the distribution of the parameter
$f$ within some interval $ (-\Delta/2,+\Delta/2)$ with $\Delta \ll
\Omega$. Then, one can estimate,

$$
(\Delta E)_f = \frac{\Delta^2}{8\Omega}.
$$

\noindent
On the other side, the number of coupled state for a fixed state from the
central band remains the same,
$
M_f \approx L/2.%
$
As a result the delocalization border can be determined from the relation,

\begin{equation}
J > J_{cr} \simeq \frac{(\Delta E)_f}{M_f} = \frac{\Delta^2}{4\Omega L}
\label{crho}
\end{equation}

This parametric dependence has been checked numerically (see
Fig.\ref{homo}) where the average number of principal component is
plotted against the rescaled interaction $J/J_{cr}$ for different
$L$ and $\Delta$. As one can see, the scaling law given by
Eq.(\ref{crho}) is quite well satisfied. Comparing with
Fig.\ref{lln}, one should note that for a constant magnetic field
the onset of a strong chaos ($N_{pc}\approx N/3$) happens in a
very small region of interaction (see the presence of small peaks
on the far right side). With further increase of the interaction,
the system again becomes nearly-integrable, since in the limit
$J\gg\Omega$ only diagonal terms dominate.

In this way we come to the same $L$-dependence for the critical
interaction $ J_{cr}$, discussed in Refs.\cite{dima0,dima}. In
these papers, the model with a nearest interaction in the plane
was considered (rather than on 1D line as in our model). For this
reason the model of Refs.\cite{dima} is free from the effects of
quasi-integrability and, therefore, the delocalization border
coincides with the border of quantum chaos.

Finally, we would like to point out that in the case of increasing
gradient of the $B^z$-magnetic field, the delocalization border
{\it increases} with an increase of the number of qubits. This
very unexpected prediction can be easily understood for the case
$\omega _k=bk^2$ (linear increase of the gradient). It can be
shown that the width $(\Delta E)_f$ grows proportional to $L^3$,
therefore, for the nearest interaction ($M_f\sim L$) the critical
interaction increases as $J_{cr}\sim L^2$, and for the
$A$-interaction one gets, $J_{cr}\sim L$. In the latter case the
estimate of $ J_{cr}$ also gives the transition to the chaos. As
one can see, the magnetic field with an increasing gradient may
strongly reduce the influence of the delocalization and chaos.

\section{Acknowledgments}

The work of GPB and VIT was supported by the Department of Energy (DOE)
under the contract W-7405-ENG-36, by the National Security Agency (NSA) and
Advanced Research and Development Activity (ARDA). FMI acknowledges the
support by CONACyT (Mexico) Grant No. 34668-E

\end{multicols}

\end{document}